\documentclass[aps,prl,twocolumn,superscriptaddress]{revtex4-1}

\usepackage{graphicx}
\usepackage{textcomp}
\usepackage{amssymb,amsmath}
\usepackage{wasysym}
\usepackage{color}
\usepackage{pdfpages}
\begin{document}

\title{A search for light dark matter in XENON10 data}

\author{J.~Angle} \affiliation{Department of Physics, University of Florida, Gainesville, FL 32611, USA} 
\affiliation{Physics Institute, University of Z\"urich, Winterthurerstrasse 190, CH-8057,  Z\"urich, Switzerland}
\author{E.~Aprile} \affiliation{Department of Physics, Columbia University, New York, NY 10027, USA}
\author{F.~Arneodo} \affiliation{Gran Sasso National Laboratory, Assergi, L'Aquila, 67010, Italy}
\author{L.~Baudis} 
\affiliation{Physics Institute, University of Z\"urich, Winterthurerstrasse 190, CH-8057,  Z\"urich, Switzerland}
\author{A.~Bernstein} \affiliation{Lawrence Livermore National Laboratory, 7000 East Ave., Livermore, CA 94550, USA}
\author{A.I.~Bolozdynya} \affiliation{Department of Physics, Case Western Reserve University, Cleveland, OH 44106, USA}
\author{L.C.C.~Coelho} \affiliation{Department of Physics, University of Coimbra, R. Larga, 3004-516, Coimbra, Portugal}
\author{C.E.~Dahl} \affiliation{Department of Physics, Princeton University, Princeton, NJ 08540, USA}
\author{L.~DeViveiros} \affiliation{Department of Physics, Brown University, Providence, RI 02912, USA}
\author{A.D.~Ferella} \affiliation{Physics Institute, University of Z\"urich, Winterthurerstrasse 190, CH-8057,  Z\"urich, Switzerland} 
\affiliation{Gran Sasso National Laboratory, Assergi, L'Aquila, 67010, Italy}
\author{L.M.P.~Fernandes} \affiliation{Department of Physics, University of Coimbra, R. Larga, 3004-516, Coimbra, Portugal}
\author{S.~Fiorucci} \affiliation{Department of Physics, Brown University, Providence, RI 02912, USA}
\author{R.J.~Gaitskell} \affiliation{Department of Physics, Brown University, Providence, RI 02912, USA}
\author{K.L.~Giboni} \affiliation{Department of Physics, Columbia University, New York, NY 10027, USA}
\author{R.~Gomez} \affiliation{Department of Physics and Astronomy, Rice University, Houston, TX 77251, USA}
\author{R.~Hasty} \affiliation{Department of Physics, Yale University, New Haven, CT 06511, USA}
\author{L.~Kastens} \affiliation{Department of Physics, Yale University, New Haven, CT 06511, USA}
\author{J.~Kwong} \affiliation{Department of Physics, Princeton University, Princeton, NJ 08540, USA}
\author{J.A.M.~Lopes} \affiliation{Department of Physics, University of Coimbra, R. Larga, 3004-516, Coimbra, Portugal}
\author{N.~Madden} \affiliation{Lawrence Livermore National Laboratory, 7000 East Ave., Livermore, CA 94550, USA}
\author{A.~Manalaysay} \affiliation{Department of Physics, University of Florida, Gainesville, FL 32611, USA}  
\affiliation{Physics Institute, University of Z\"urich, Winterthurerstrasse 190, CH-8057,  Z\"urich, Switzerland}
\author{A.~Manzur} 
\affiliation{Department of Physics, Yale University, New Haven, CT 06511, USA}
\author{D.N.~McKinsey} \affiliation{Department of Physics, Yale University, New Haven, CT 06511, USA}
\author{M.E.~Monzani} \affiliation{Department of Physics, Columbia University, New York, NY 10027, USA}
\author{K.~Ni} \affiliation{Department of Physics, Yale University, New Haven, CT 06511, USA}
\author{U.~Oberlack} \affiliation{Department of Physics and Astronomy, Rice University, Houston, TX 77251, USA} \affiliation{Johannes Gutenberg University Mainz, 55099 Mainz,Germany}
\author{J.~Orboeck} \affiliation{Department of Physics, RWTH Aachen University, Aachen, 52074, Germany}
\author{G.~Plante} \affiliation{Department of Physics, Columbia University, New York, NY 10027, USA}
\author{R.~Santorelli} \affiliation{Department of Physics, Columbia University, New York, NY 10027, USA}
\author{J.M.F.~dos~Santos} \affiliation{Department of Physics, University of Coimbra, R. Larga, 3004-516, Coimbra, Portugal}
\author{S.~Schulte} \affiliation{Department of Physics, RWTH Aachen University, Aachen, 52074, Germany}
\author{P.~Shagin} \affiliation{Department of Physics and Astronomy, Rice University, Houston, TX 77251, USA}
\author{T.~Shutt} \affiliation{Department of Physics, Case Western Reserve University, Cleveland, OH 44106, USA}
\author{P.~Sorensen}
\email{pfs@llnl.gov}
\affiliation{Lawrence Livermore National Laboratory, 7000 East Ave., Livermore, CA 94550, USA}
\author{C.~Winant} \affiliation{Lawrence Livermore National Laboratory, 7000 East Ave., Livermore, CA 94550, USA}
\author{M.~Yamashita} 
\affiliation{Department of Physics, Columbia University, New York, NY 10027, USA}

\collaboration{XENON10 Collaboration}\noaffiliation
\date{June 3, 2011}
\begin{abstract}
We report results of a search for light ($\lesssim10$~GeV) particle dark matter with the XENON10 detector.  The event trigger was sensitive to a single electron, with the analysis threshold of 5 electrons corresponding to 1.4~keV nuclear recoil energy.  Considering spin-independent dark matter-nucleon scattering, we exclude cross sections $\sigma_n>7\times10^{-42}$~cm$^2$, for a dark matter particle mass $m_{\chi}=7$~GeV.  We find that our data strongly constrain recent elastic dark matter interpretations of excess low-energy events observed by CoGeNT and CRESST-II, as well as the DAMA annual modulation signal.
\end{abstract}

\maketitle

Recently, the CRESST-II and CoGeNT collaborations have reported the observation of low-energy events in excess of known backgrounds \cite{2010seidel, 2010aalseth}.  This has encouraged the hypothesis that these signals $-$ in addition to the long-standing DAMA \cite{2008bernabei} annual modulation signal $-$ might arise from the scattering of a light ($\lesssim10$~GeV) dark matter particle \cite{2010chang3, 2010graham,2010essig, 2010hooper, 2010fitzpatrick,2010feldstein,2009bottino}.  The CDMS \cite{2010akerib, 2010ahmed} dark matter search data been have re-analyzed with lowered energy thresholds, but do not fully exclude light dark matter interpretations.  In order to maintain sensitivity to the scattering of such light galactic dark matter, an experiment needs either a very low $\mathcal{O}$(keV) energy threshold, as with CoGeNT, or light target nuclei, such as the oxygen atoms in the CRESST-II detector.  This is a simple requirement of the kinematics \cite{1996lewin}.  For example, consider a halo-bound dark matter particle with mass $m_{\chi}=10$~GeV and velocity 600~km~s$^{-1}$.  The maximum recoil energy that would result from an elastic scatter of such a particle in an earth-bound target would be about 20~keV for a recoiling oxygen nucleus (CRESST-II), 9~keV for germanium (CoGeNT) and only 6~keV for xenon.  The respective energy thresholds of CRESST-II and CoGeNT are approximately 10~keV \cite{2009angloher} and $2$~keV \cite{2010aalseth}, which combined with a low background event rate results in good sensitivity to light mass dark matter.  The energy threshold of previously reported XENON10 data \cite{2009angle,2008angle2} depended on the primary scintillation efficiency of liquid xenon for nuclear recoils ($\mathcal{L}_{eff}$) \cite{2010manzur,2009aprile}.  For a conservative assumption of the energy dependence of $\mathcal{L}_{eff}$ \cite{2010manzur}, the threshold was about 5~keV.

It is possible to obtain a lower energy threshold from existing XENON10 dark matter search data, if nuclear recoil energy is measured by the detected electron signal.  The method allows us to reach an energy threshold E$_{nr}\sim1$~keV.  At such low nuclear recoil energies, the primary scintillation signal is generally absent.  As a result, two important aspects of the XENON10 detector \cite{2010aprile} performance are compromised:  the ability to precisely reconstruct the $z$ coordinate of a particle interaction, and the discrimination between incident particle types.  The detected ratio of scintillation to electron signals was used in \cite{2009angle,2008angle2} to discriminate and reject about 99.5\% of electromagnetic background events which would otherwise have been treated as dark matter candidate events.  Loss of this discrimination thus reduces our sensitivity to nuclear recoils from dark matter scattering, by about two orders of magnitude.  Still, the lower energy threshold we obtain permits the exploration of new regions of dark matter particle mass ($m_{\chi}$) and cross section ($\sigma_n$) parameter space.

The XENON10 detector is described in detail in \cite{2010aprile}.  For the present discussion, we remind the reader that XENON10 observes particle interactions via detected primary scintillation photons (the S1 signal) and electrons (the S2 signal).  The electrons are drifted across the liquid xenon target and extracted into gaseous xenon, where they create proportional scintillation photons.  This allows both S1 and S2 to be observed with the same photomultiplier tube arrays.   A single electron extracted from the liquid target results in about 27~photoelectrons in the photomultiplier arrays.  It is this robust signal that gives the S2 channel its lower energy threshold. The $(x,y)$ coordinates of particle interactions are reconstructed from the hit pattern of the S2 signal on the top photomultipliers.  The $z$ coordinate is usually reconstructed from $z = v_d \Delta t$, where $v_d\simeq0.20$~cm~$\mu$s$^{-1}$ is the electron drift velocity \cite{1982gushchin} and $\Delta t$ is the measured time delay between the S1 and S2.  This method cannot be used if there is no S1 signal, as is often the case for very low energy nuclear recoils \cite{2009sorensen}.  The S2 pulse width $\sigma_e$ carries a mild $z$-dependence \cite{2010sorensen_idm}.  Larger detectors \cite{2010aprile100,2010mckinsey}, if operated with a lower value of $E_d$, may obtain a reliable estimate of the $z$ coordinate from $\sigma_e$ \cite{2011sorensen_nim}; this should lead to a significant rejection of edge (in $z$) events and a commensurate improvement in sensitivity. 

\begin{figure}[h]
\centering
\includegraphics[width=0.45\textwidth]{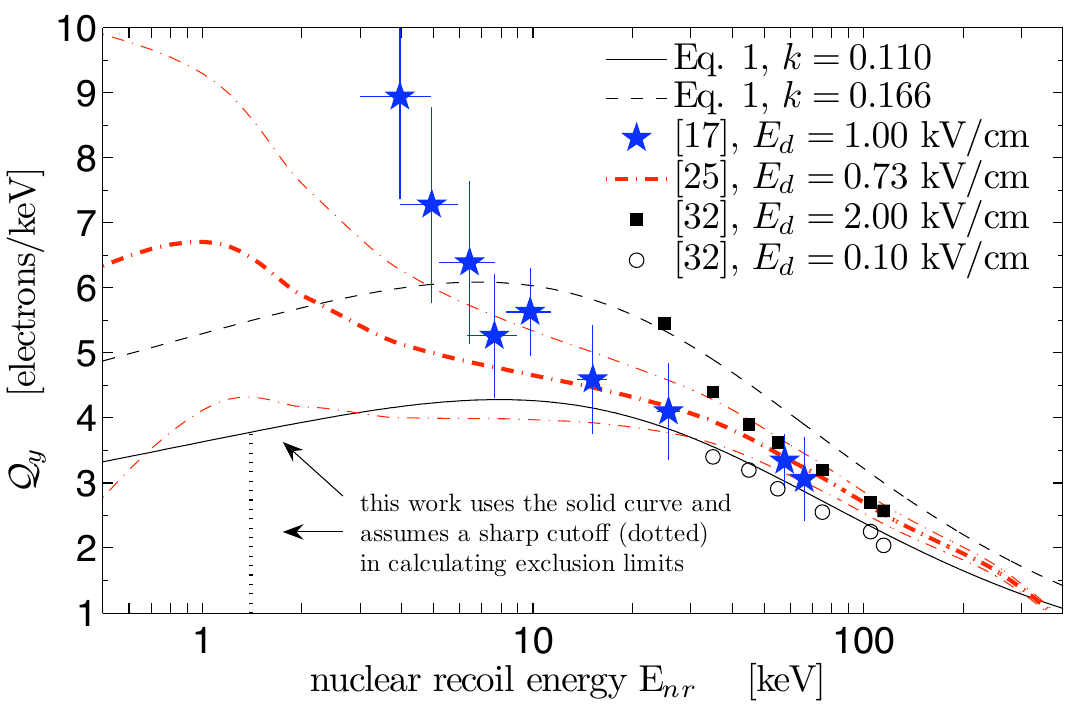}
\caption{The electron yield $\mathcal{Q}_y$ of liquid xenon for nuclear recoils.  Theoretical curves (solid and dashed) were calculated from Eq. \ref{eq1}, as described in the text.  Also shown are measurements from \cite{2010manzur} ($\bigstar$), \cite{2010sorensen_idm} (dash-dot curve, with $\pm1\sigma$ contours) and \cite{2006aprile} ($\fullmoon$ and $\blacksquare$, uncertainty omitted for clarity). }
\label{fig1}
\end{figure}

Ideally we would like to reconstruct the nuclear recoil energy for each event from $\mbox{E}_{nr} = \epsilon(n_{\gamma} + n_e)/f_n$, as in \cite{2011sorensen,2007shutt}, with $\epsilon=13.8$~eV the average energy to create a photon or electron, and $f_n$ the nuclear recoil quenching \cite{1963lindhard}.   However, at low recoil energies the small number $n_{\gamma}$ of primary scintillation photons often does not result in a measurable S1 response.  Since we are interested in events at very low recoil energy, we calibrate the energy scale using only the number $n_e$ of measured electrons in the S2 signal.  The electron yield of liquid xenon for nuclear recoils has been measured directly using tagged neutron scattering \cite{2010manzur,2006aprile}.  The lowest-energy data point from \cite{2010manzur} implies an S2 signal of $36\pm6$~electrons in the $3-5$~keV range, as shown in Fig. \ref{fig1}.  An indirect measurement of the electron yield is described in \cite{2010sorensen_idm}, and was obtained following the method detailed in \cite{2009sorensen}.  The central and $\pm1\sigma$ contours of that work are shown in Fig. \ref{fig1} (dash-dot curves).  That $\mathcal{Q}_y$ rises with decreasing E$_{nr}$ between 100 and 10~keV is a result of the increasing fraction of nuclear recoil energy given to electrons (rather than photons) over that energy range \cite{2011sorensen}.

Following \cite{2011sorensen}, we obtain a theoretical prediction for the electron yield, 
\begin{equation}\label{eq1}
\mathcal{Q}_y \equiv \frac{n_e}{\mbox{E}_{nr}} = \frac{1}{\xi} \mbox{ln}(1+\xi) \frac{f_n(k)/\epsilon}{1+N_{ex}/N_i}.
\end{equation}
In this equation, $N_{ex}/N_i$ is the number ratio of excited to ionized xenon atoms, and $\xi=N_i \alpha /(4a^2 v)$ is the single parameter upon which the Thomas-Imel box model depends \cite{1987thomas,foot1}.  In Eq. \ref{eq1} we explicitly indicate the dependence of $f_n$ on the proportionality constant $k$, between the velocity of a xenon nucleus and its electronic stopping power.  In Fig. \ref{fig1} we show Eq. \ref{eq1} predictions for two $k$ values, which correspond to calculations by Lindhard \cite{1963lindhard} and Hitachi \cite{2006aprile}.  The dashed curve is the best-fit case from \cite{2011sorensen}, with $k=0.166$, $N_{ex}/N_i=1.05$ and $4\xi/N_i=0.024$.  The solid curve, which we will use to calibrate the energy scale in the present work, takes the more conservative $k=0.110$, from which we obtain the best-fit parameters $N_{ex}/N_i=1.09$ and $4\xi/N_i=0.032$.  This results in the most conservative $\sigma_n$ exclusion limits based on available data and theoretical considerations, and is consistent with our neutron calibration data \cite{2010sorensen_idm}.  However, it is in tension with the measurements of Ref. \cite{2010manzur} below $\sim7$~keV.  As discussed in \cite{2010sorensen}, the rising measured $\mathcal{Q}_y$ values in this regime could be influenced by trigger threshold bias. We emphasize that our energy calibration (Fig. \ref{fig1}, solid curve) relies on extrapolation of Lindhard's theory \cite{1963lindhard} below 4~keV.

We report results from a 12.5 live day dark matter search, obtained between August 23 and September 14, 2006.  This data set is distinct from the previously reported \cite{2009angle,2008angle2} XENON10 dark matter search data:  the present data was obtained with the the secondary scintillation gain about 12\% higher, and the S2-sensitive trigger threshold set at the level of a single electron.  The trigger efficiency for single electrons is $>0.80$ \cite{2008sorensen}.  

Event selection criteria are summarized in Table \ref{table1}.  Candidate events were required to have an S2 signal of at least 5 electrons, or 1.4 keV.  Although the detector is sensitive to smaller energy depositions, we are unable to reliably assess the cut acceptance $\varepsilon_c$ for smaller nuclear recoils, due to the trigger configuration during the neutron calibration \cite{2010aprile}.  The position of interaction was required to fall within $r<3$~cm.  This central region features optimal self-shielding by the surrounding xenon target.  A signal-to-noise cut required the S2 pulse to contain at least 0.45 of the total area of the event record.  The acceptance of this cut rises monotonically from 0.94 to $>0.99$ between 1.4~keV and 10~keV.  Valid single scatter event records were required to have only a single S2 pulse of size $>4$~electrons.  Events in which an S1 signal was found were required to have log$_{10}$(S2/S1) within the $\pm3\sigma$ band for elastic single scatter nuclear recoils.  This band was determined from the neutron calibration data, and has been reported in a previous article \cite{2009angle}.  Events in which no S1 signal was found were assumed to be dark matter candidate events and were retained.

\begin{table}[h]
\centering
\caption{Summary of cuts applied to 15 kg-days of dark matter search data, corresponding acceptance for nuclear recoils $\varepsilon_c$ and number of events remaining in the range $1.4<\mbox{E}_{nr}\leq10$~keV.}
\begin{tabular}{llrr}
\\
\toprule
~& Cut description & $\varepsilon_c$~~~~ & $\mbox{N}_{evts}$ \\
\hline
1. & event localization $r<3$~cm & 1.00$^a$  & 125 \\ 
2. & signal-to-noise & $>0.94$~  & 58\\ 
3. & single scatter (single S2) & $>0.99$~  & 38\\ 
4. & $\pm3\sigma$ nuclear recoil band & $>0.99$~ & 23\\
\botrule
\multicolumn{3}{l}{$^a$ limits effective target mass to 1.2 kg}\\
\end{tabular}
\label{table1}
\end{table}

\begin{figure}[t]
\centering
\includegraphics[width=0.45\textwidth]{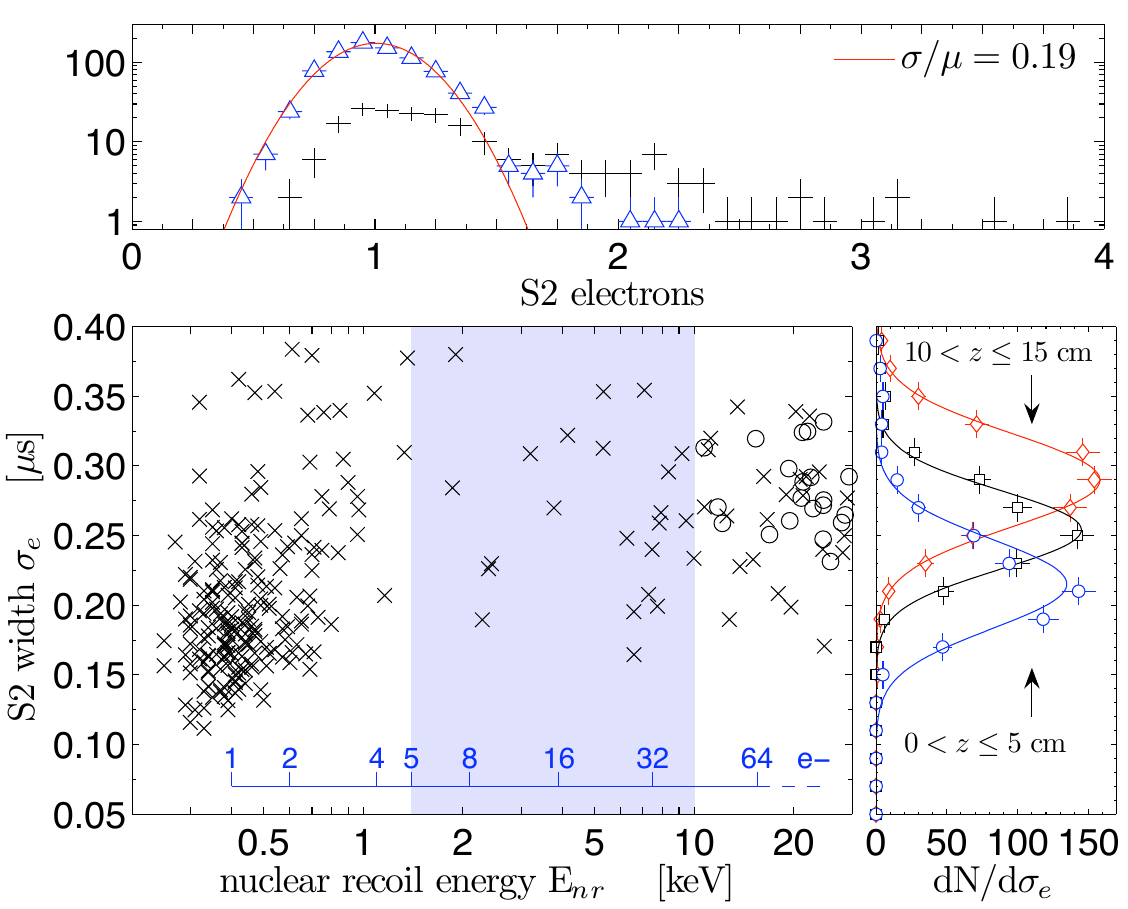}
\caption{{\bf (left)} All candidate dark matter events remaining ($\times$ and $\fullmoon$) after the cuts listed in Table \ref{table1}.  Events in which an S1 was found are shown as $\fullmoon$.  The number of electrons in the S2 signal is indicated by the inset scale. {\bf (top)} Distribution of candidate events with $\leq4$~electrons ($+$), and distribution of background single electrons ($\triangle$) as described in the text. {\bf (right)} S2 pulse width distributions for single scatter nuclear recoils in the top, middle and bottom third of the detector.}
\label{fig2}
\end{figure}

The remaining events in the lowest-energy region are shown in Fig. \ref{fig2} (left) versus their S2 pulse width $\sigma_e$. The equivalent number of electrons is indicated by the inset scale.  Events in which an S1 signal was observed are indicated by a circle.  Figure \ref{fig2} (right) shows the width profile of the S2 signal in the top, middle and bottom third of the detector, based on single scatter nuclear recoils with known $\Delta t$ and $5<\mbox{S2}<100$ electrons.  Gaussian fits are shown to guide the eye. 

The top panel of Fig. \ref{fig2} shows the distribution of remaining candidate events ($+$) with S2~$\leq4$~electrons.  The distribution of background single electron events, sampled from a time window at least $20~\mu$s after higher-energy events, is also shown ($\triangle$).  The single electron background events are a subject of ongoing study, and appear to originate from multiple physical phenomena.Ê One possibility involves photoionization of impurities in the liquid xenon \cite{2007edwards}.Ê Another possible origin is from excess free electrons trapped at the liquid surface.Ê This could occur because the emission of electrons from the liquid to the gas is nearly $-$ but likely not exactly $-$ unity \cite{1982gushchin1}.Ê As a result, every S2 signal could be a potential source of a small number of trapped electrons. Delayed emission of the trapped electrons may result from the requirement that both the electron kinetic energy and the $z$ component of the electron momentum be sufficient to overcome the surface potential barrier \cite{1999bolozdynya}.  

The signal-to-noise cut was motivated by a distinct but closely related class of background event, which consists of a train of approximately ten to several tens of single electrons over a period of $\mathcal{O}(100~\mu\mbox{s})$.  The origin of these events is also not yet clear.  Often several single electrons in an electron train overlap in time, to the degree that they appear as a single S2 pulse containing $\sim2-6$ electrons.  These spurious pulses often have $\sigma_e>0.30$ (the $3\sigma$ width for a single electron) and so could be removed based on pulse width.  However, the signal-to-noise cut more precisely targets the presence of multiple additional single electrons in the event record.

\begin{figure}[h]
\centering
\includegraphics[width=0.45\textwidth]{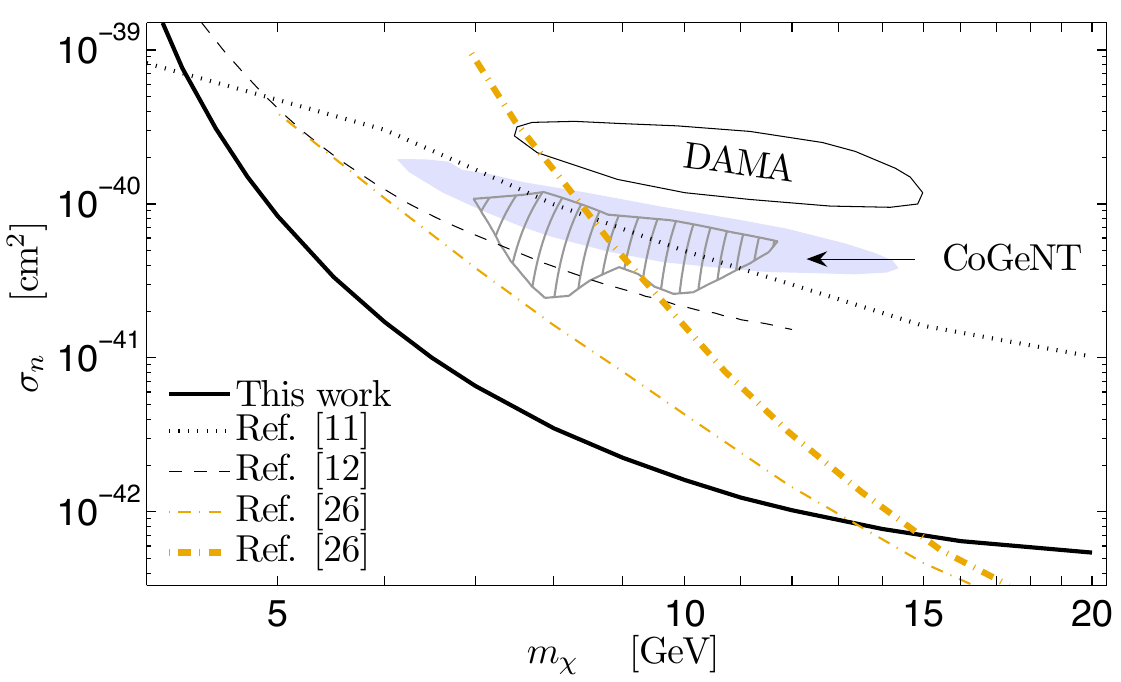}
\caption{Curves indicate 90\% C.L. exclusion limits on spin-independent $\sigma_n$ for elastic dark matter scattering, obtained by CDMS (dotted \cite{2010akerib}, and dashed \cite{2010ahmed}) and XENON100 (dash-dot \cite{2010aprile100}).  The region consistent with assumption of a positive detection by CoGeNT is shown (hatched) \cite{2010aalseth}, and (shaded) \cite{2010chang3}; the latter assumes a 30\% exponential background.   Also shown is the $3\sigma$ allowed region for the DAMA annual modulation signal (solid contour) \cite{2011savage}. } 
\label{fig3}
\end{figure}

The energy resolution for S2 signals depends primarily on Poisson fluctuation in the number of detected electrons, with an additional component due to instrumental fluctuations.  This is discussed in detail in \cite{2010sorensen}, and for higher energy signals in \cite{2010aprile}.   So as not to overstate the energy resolution, we adopt a parameterization which follows the Poisson component only, given by $\mathcal{R}(\mbox{E}_{nr})=(2\mbox{E}_{nr})^{-1/2}$.  We assume a sharp cutoff in $\mathcal{Q}_y$ at E$_{nr}=1.4$~keV, and then convolve the resolution with the predicted differential dark matter scattering rate.  This ensures that $\sigma_n$ exclusion limits are not influenced by lower-energy extrapolation of the detector response.  The scattering rate as a function of nuclear recoil energy was calculated in the usual manner \cite{1996lewin} ({\it cf.} \cite{2009angle}).  We take the rotational speed of the local standard of rest and the velocity dispersion of the dark matter halo to be $v_0=230$~km~s$^{-1}$, and the galactic escape velocity to be $v_{esc}=600$~km~s$^{-1}$ \cite{2007smith}.  We use the $p_{{max}}$ method \cite{2002yellin} to calculate 90\% C.L. exclusion limits on the cross section $\sigma_n$ for elastic spin-independent dark matter $-$ nucleon scattering as a function of $m_{\chi}$.  All remaining events in the the range $\mbox{E}_{nr}>1.4$~keV are treated as potential dark matter signal.  The results are shown in Fig. \ref{fig3}.  If $\mathcal{Q}_y$ were 40\% higher (lower) below 4 keV, the exclusion limits would be about $\times2$ stronger (weaker) at $m_{\chi}=7$~GeV.

The exclusion limits and allowed regions shown in Fig. \ref{fig3} assume a simple Maxwell-Boltzman distribution for the dark matter halo.  Given the likelihood of significant departures from this distribution \cite{2010kuhlen}, it is important to understand if astrophysical uncertainties could alter the incompatibility of our results with the positive detection scenarios shown in Fig. \ref{fig3}.  A method for doing so is described in \cite{2010fox}, and predicts that not less than $\sim5$~counts~keV$^{-1}$~kg$^{-1}$~day$^{-1}$ should be observed in a xenon detector, if the unexplained low-energy rise observed by the CoGeNT detector \cite{2010aalseth} were due to dark matter scattering.  It can be seen from Table \ref{table1} that we observe an event rate of $\sim0.2$~counts~keV$^{-1}$~kg$^{-1}$~day$^{-1}$ on the interval $1.4<\mbox{E}_{nr}<10$~keV, indicating that the order of magnitude exclusion of the CoGeNT regions shown in Fig. \ref{fig3} is robust against astrophysical uncertainties.  Due to the preliminary nature of the CRESST-II results we do not show a corresponding allowed region, although it appears likely to lie above the DAMA region, as shown in Fig. \ref{fig3} of Ref. \cite{2010schwetz}. 

We have shown for the first time that it is possible to perform a sensitive search for dark matter with a liquid xenon time-projection chamber, using only the electron signal. The advantage of this analysis is an increased sensitivity to light ($\lesssim10$~GeV) dark matter candidate particles, due to the approximate factor $\times5$ decrease in the detector energy threshold.  For larger particle masses, standard analyses \cite{2009angle,2010aprile100,2011aprile} offer superior sensitivity.  The present work appears to severely constrain recent light elastic dark matter interpretations of the excess low-energy events observed by CoGeNT and CRESST-II, as well as interpretations of the DAMA modulation signal.

This work was initiated at the KITP workshop ``Direct, Indirect and Collider Signals of Dark Matter,'' Santa Barbara CA, December 7-18, 2009, which was supported in part by the National Science Foundation under Grant No. PHY-05-51164.  We gratefully acknowledge support from NSF Grants No. PHY-03-02646 and No. PHY-04-00596, CAREER Grant No. PHY-0542066, DOE Grant No. DE-FG02-91ER40688, NIH Grant No. RR19895, SNF Grant No. 20-118119, FCT Grant No. POCI/FIS/60534/2004 and the Volskwagen Foundation. 


\clearpage
\includepdf{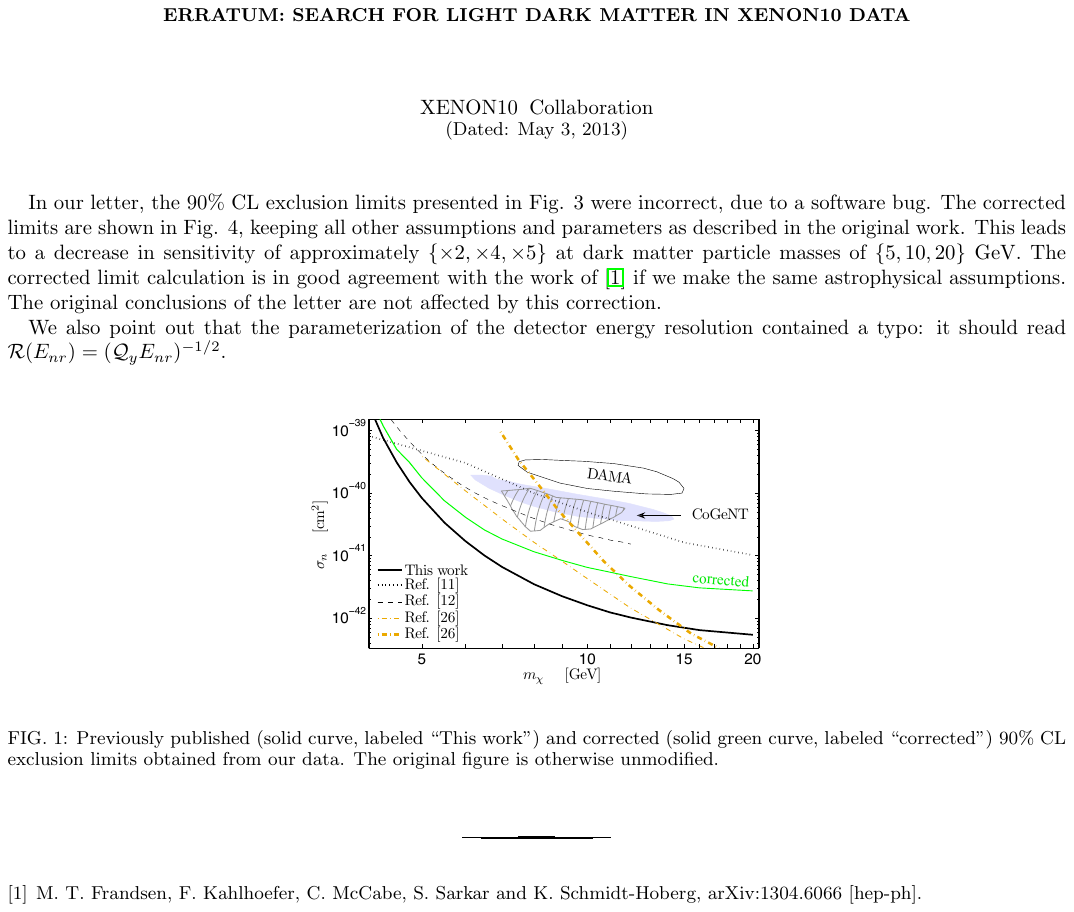}

\end{document}